\documentclass[onecolumn,aps,prd,showpacs,superscriptaddress,nofootinbib,amsmath,amssymb,floats,floatfix,showkeys,notitlepage,longbibliography]{revtex4-1}
\usepackage{comment}
\usepackage{xcolor}
\usepackage{graphicx}
\usepackage{subfigure}
\usepackage{palatino}
\usepackage[commandnameprefix=always]{changes}
\usepackage{hyperref}
\hypersetup{colorlinks=true,linkcolor=blue,urlcolor=blue,citecolor=blue}
\usepackage[toc,page]{appendix}
\usepackage[normalem]{ulem}
\usepackage{orcidlink}
\usepackage{lipsum}
\usepackage{graphicx}
\usepackage{subfigure}
\usepackage{palatino}
\usepackage{sans}
\usepackage{adjustbox}
\usepackage{latexsym}
\usepackage{amsmath}
\usepackage{amssymb}
\usepackage{amsfonts}
\usepackage{dcolumn}
\usepackage{bm}
\usepackage{tikz}
\usepackage{bigints}
\usepackage{array,tabularx,multirow,booktabs}
\usepackage[tracking=true]{microtype}
\SetTracking{}{500}
\SetTracking{encoding={*}, shape=sc}{40}
\UseRawInputEncoding %for inputenc error%
\allowdisplaybreaks
\usepackage{adjustbox}
\usepackage{latexsym}
\usepackage{amsmath}
\usepackage{amssymb}
\usepackage{amsfonts}
\usepackage{dcolumn}
\usepackage{bm}
\usepackage{tikz}
\usepackage{bigints}
\usepackage{array,tabularx,multirow,booktabs}
\usepackage[tracking=true]{microtype}
\usepackage{color}
\UseRawInputEncoding %for inputenc error%
\allowdisplaybreaks

\begin{document} 

\title{Regular Black Hole from gravitational collapse of dust and radiation}

\author{Vitalii Vertogradov}
\email{vdvertogradov@gmail.com}
\affiliation{Physics department, Herzen state Pedagogical University of Russia,
48 Moika Emb., Saint Petersburg 191186, Russia} 
\affiliation{SPB branch of SAO RAS, 65 Pulkovskoe Rd, Saint Petersburg
196140, Russia}

\begin{abstract}
A black hole is the end state of the gravitational collapse of massive stars. However, a typical black hole contains a singularity and to avoid singularity formation we have to violate a strong energy condition that states that gravity must attract. The nature of the matter that prevents the singularity formation is still unknown.  In this paper, we offer a simple model of gravitational collapse of dust and radiation. In the simplest case, such a model leads to a singular black hole. However, if we assume an energy exchange between dust and radiation, which increases towards the center of the collapsing cloud, then it becomes possible to construct a model of a regular black hole. We identify the conditions under which a regular center is possible and also give an explicit example of how such an interaction leads to a regular center.

\end{abstract}

\date{\today}



\maketitle
%\tableofcontents
\section{Introduction}
In 1916 Karl Schwarzschild obtained the first solution of Einstein field  equations, which, as it turned out later, describes a black hole. This solution has two singularities at the event horizon and at the center. While the singularity at the event horizon is purely coordinate and can be removed by a suitable choice of coordinates, the singularity at the center is true, i.e. it cannot be removed by any coordinate transformation. Despite the fact that black holes are formed in the process of gravitational collapse of massive stars, they remained exotic and purely theoretical objects for a long time. However, that changed after the first images of a black hole shadow were obtained by the Event Horizon Telescope Collaboration. In 2019, a black hole located at the center of the M87 galaxy was observed~\cite{bib:eht1}, and three years later, they obtained images  of a supermassive black hole Sagittarius $A^*$, located at the center of our Milky Way galaxy~\cite{bib:eht2}. These observations moved black holes from the class of exotic theoretical objects to the class of real astrophysical ones. 
However, the singularity at the center of the black hole indicates that the general relativity is not proper theory to describe processes in the vicinity of this region. Moreover, the famous Penrose's theorem on singularities~\cite{bib:pen} states that during gravitational collapse, when the apparent horizon  forms, a singularity is also inevitably formed.
Recently regular black holes - black hole in the center of which singularity is replaced by a de Sitter core, have become very popular~\cite{bib:ansoldi, bib:review, bib:Bonanno2023rzk, bib:bronnikov, bib:khlopov, bib:khlopov2}. The idea that superdense matter passes into a vacuum medium belongs to Gliner~\cite{bib:gliner} and Sakharov~\cite{bib:sakharov}. Bardeen was the first who constructed the model of a regular black hole~\cite{bib:bardeen}, and it was later found that this solution was supported by nonlinear electrodynamics. After that, other important models of regular black holes were found such as Hayward black hole~\cite{bib:hay}, Dymnikova black hole~\cite{bib:dym} and others. Recently, it has been found that the Hagedorn fluid can also be the source of a regular black hole~\cite{bib:ali2025cqg}. 
However, a black hole is formed as a result of the gravitational collapse of an usual massive star, inside which there is no such exotic matter. Oppenheimer and Sneider were the first who offered the model of gravitational collapse of dust ball leading to black hole formation~\cite{bib:open}. Later it was found out that gravitational collapse can also lead to the naked singularity formation~\cite{bib:joshi_book, bib:joshi_review, bib:vertogradov2024grg, bib:maombi}. But it is still not clear how the collapse of ordinary matter transitions into the exotic matter which is necessary for a regular black hole formation. 
The aim of this paper is to construct a dynamical model and to show that the collapse of ordinary matter can lead to the formation of a regular center.  We have considered the collapse of a matter cloud consisting of dust and radiation. We also assumed that in the process of gravitational collapse the matter passes to radiation and the denser the object becomes, the effect increases. It also increases to the center of collapsing cloud. This means in particular that in spite of the fact that the total energy-momentum tensor is conserved, its separate parts corresponding to dust and radiation do not satisfy the conservation law. Under such assumptions it is possible to construct a model which describes a regular black hole. 

this paper is organized as follows: in sec. 2 we construct the model of gravitational collapse of a cloud consisting of dust and radiation without any interaction. In sec. 3 the same model is considered but additionaly we assume that there is interaction. In sec. 4 we show that if we consider the model of interaction between dust and radiation which increases to the center, then the regular black hole formation is possible. Section 5 is discussion of obtained results.
Throughout the paper the geometrized system of units $c=8\pi G=1$ is adopted. We also use the signature $-+++$.
\section{A model without interaction}
We begin our consideration from the simplest case when matter and radiation do not interact. For this purpose we assume the most general form of a dynamical spherically symmetric spacetime written in Eddington-Finkelstein coordinates $\{v, r, \theta, \varphi\}$ as
\begin{equation} \label{eq:metric}
ds^2=-\left(1-\frac{2M(v,r)}{r}\right)dv^2+2\varepsilon dvdr +r^2d\Omega^2,
\end{equation}
here $M(v,r)$ is the mass function of both radial coordinate $r$ and Eddington time $v$, $\varepsilon =\pm 1$ depending on ingoing or outgoing energy flux and $d\Omega^2=d\theta^2+\sin^2 \theta d\varphi^2$ is a metric on unit two-sphere. Without loss of generality, we assume $\varepsilon =+1$, i.e. we are mostly interested in black hole formation. 

The physical quantities for the spacetime \eqref{eq:metric} are given by
\begin{eqnarray} \label{eq:density}
\sigma (v,r)&=&\frac{2\dot{M}(v,r)}{r^2},\nonumber \\
\rho (v,r)&=&\frac{2M'(v,r)}{r^2},\nonumber \\
P(v,r)&=&-\frac{M''}{r}.
\end{eqnarray}
Where $\sigma$ is associated with density of energy flux and $P$ and $\rho$ are pressure and energy density of the matter respectively.
We assume that energy density and pressure are connected through barotropic equation of state
\begin{equation}
P=\alpha \rho.
\end{equation}
Now, we assume that the collapsing cloud consists of two components - dust and radiation. Thus the total energy density will be written in the form
\begin{equation}
\rho=\rho_d+\rho_r.
\end{equation}
The pressures will also be a combination of dust and radiation, i.e.
\begin{equation}
P=P_d+P_r.
\end{equation}
As the result we have two equations of state
\begin{equation}
P_d=0,~~ P_r=\frac{1}{3}\rho_r.
\end{equation}
One of the continuity equations follows from the conservation of the energy-momentum tensor $T^{ik}_{;k}=0$ and it is given by
\begin{equation} \label{eq:conservation}
\rho'r+2\rho+2P=0.
\end{equation}
As the first model, we consider the model in which besides the conservation of the total energy-momentum tensor, also its separate parts are conserved. It means that in the process of gravitational collapse there is no energy exchange from matter to radiation or vice versa. This can be written as

\begin{eqnarray}
\rho'_dr&+&2\rho_d=0,\nonumber \\
\rho'_rr&+&\frac{8}{3}\rho_r=0.
\end{eqnarray}
Solving each equation separately we find the dust and radiation densities as functions of coordinate $r$ and time$v$.
\begin{eqnarray}
\rho_d(v,r)&=&\rho_{0d}(v)r^{-2},\nonumber \\
\rho_r(v,r)&=&\rho_{0r}(v)r^{-\frac{8}{3}}.
\end{eqnarray}
By using \eqref{eq:density}, i.e.
\begin{equation}
\rho=\rho_{0d}r^{-2}+\rho_{0r}r^{-\frac{8}{3}}=\frac{2M'}{r^2},
\end{equation}
we come to the mass function of the form
\begin{equation} \label{eq:mass1}
M(v,r)=M_0(v)+\frac{1}{2}\rho_{0d}r+\frac{3}{2}\rho_{0r}r^{\frac{1}{3}}.
\end{equation}
The solution \eqref{eq:mass1} is the combination of Husain solution for dust and radiation~\cite{bib:husain1998exact, bib:vertogradov2016gc}.
This solution describes a black hole and has at least one apparent horizon. In order to find out whether this solution is regular or not, it is necessary to calculate the curvature invariants , i.e., the Ricci scalar $R$, the square Ricci tensor  $S=R_{ik}R^{ik}$ and the Kretschmann scalar $K=R_{iklm}R^{iklm}$, and to make sure that they do not diverge anywhere in spacetime. These carvature invariants, in the general case for a metric \eqref{eq:metric}, are given by 
\begin{eqnarray} \label{eq:carvature}
R&=&\frac{4M'+2rM''}{r^2},\nonumber \\
S&=&\frac{8M'^2+2 r^2M''^2}{r^4},\nonumber \\
K&=&\frac{48M^2-64rMM'+32r^2M'^2+16r^2MM''-16r^3M'M''+4r^4M''^2}{r^6}.
\end{eqnarray}
One can rewrite these expressions through the mass function $M(v,r)$, energy density $\rho$ and pressure $P$ as
\begin{eqnarray} \label{eq:carvature2}
R&=&2\rho-2P,\nonumber \\
S&=&2\rho^2+2P^2 \\
K&=&\frac{48M^2}{r^6}-\frac{16M}{r^3}\left(2\rho-P\right)+8\rho^2-8\rho P+4P^2.
\end{eqnarray}
For the curvature invariants to have a finite value at the center  at $r\rightarrow 0$, the following three conditions must be satisfied
\begin{itemize}
\item $\lim\limits_{r\rightarrow 0}M(r)=0$;
\item $\lim\limits_{r \rightarrow 0} \rho (r)=\rho_0=const.$;
\item $\lim\limits_{r\rightarrow 0} P(r)=P_0=const.$.
\end{itemize}
Obviously, the solution  \eqref{eq:mass1} leads to singularity formation in the center. since none of the above conditions is satisfied: the mass function does not go to zero
\begin{equation}
\lim\limits_{r\rightarrow 0} M(v,r)=M_0(v)\neq 0,
\end{equation}
 and the energy density and pressure 
\begin{eqnarray}
&&\lim\limits_{r\rightarrow 0}\frac{\rho_{0d}(v)}{r^2}+\frac{\rho_{0r}(v)}{r^{\frac{8}{3}}} \rightarrow +\infty,\nonumber \\
&&\lim\limits_{r\rightarrow 0}\frac{\rho_{0r}(v)}{3r^{\frac{8}{3}}}\rightarrow +\infty,
\end{eqnarray}
diverge.

\section{A model with interaction}
Now we assume that the momentum energy tensor is conserved, but its individual parts are not conserved and there is a transfer of energy from one component of matter to another, i.e.
\begin{eqnarray}
\rho'_dr&+&2\rho_d=-\beta \rho_r,\nonumber \\
\rho'_rr&+&\frac{8}{3}\rho_r=\beta \rho_r.
\end{eqnarray}
Here $\beta$ is dimensionless constant. If it is positive then there is energy transfer from dust to radiation and if it is negative then transfer is in opposite direction.
By solving this system of differential equations, one obtains
\begin{eqnarray}
\rho_r&=&\rho_{0r}r^{\beta-\frac{8}{3}},\nonumber \\
\rho_m&=&r^{-2}\left(\rho_{0m}-\frac{\beta \rho_{0r}}{\beta-\frac{2}{3}}r^{\beta-\frac{2}{3}}\right).
\end{eqnarray}
And the mass function then is given by
\begin{equation} \label{eq:mass2}
M(v,r)=M_0(v)+\frac{1}{2}\rho_{0m}(v)r-\frac{\rho_{0r}}{3(\beta -\frac{2}{3})(\beta+\frac{1}{3})}r^{\beta+\frac{1}{3}}.
\end{equation}
Note that if $\beta \rightarrow 0$ then solution \eqref{eq:mass2} becomes \eqref{eq:mass1}. One important remark is that if we take $\beta\geq \frac{8}{3}$ then it becomes possible to obtain finite energy density and pressure in the centre if the following condition is held
\begin{equation}
\rho_{0m}(v)=\frac{\beta \rho_{0r}(v)}{\beta-\frac{2}{3}}.
\end{equation}
However, the solution \eqref{eq:mass2} doesn't lead to regular black hole formation because one can't satisfy the condition $\lim\limits_{r\rightarrow 0}M(v,r)=0$. This limit always leads to gravitationally weak singularity - important property because the solution \eqref{eq:mass1} always lead to gravitationally strong singularity~\cite{bib:tip, bib:ir, bib:clorak}. One should also note, that when gravitational collapse starts  at $v=0$ singularity is apsent and the region near this regular core might be visible by faraway observer.
To prove this statement let us consider radial null geodesic which is given by
\begin{equation}
\frac{dv}{dr}=\frac{2}{1-\frac{2M(v,r)}{r}}.
\end{equation}
If this geodesic terminates in the centre in the past and it is future-directed then we can obtain information from this region. It might happen if the quantity $X_0$ defined as
\begin{equation} \label{eq:defx}
\lim\limits_{v\rightarrow 0, r\rightarrow 0}\frac{dv}{dr}=X_0,
\end{equation}
has positive and finite value. Let us substitute to the geodesic equation the mass function \eqref{eq:mass2} and let us assume for simplisity that
\begin{eqnarray}
M_0(v)&=&\mu v,~~ \mu>0,\nonumber \\
\rho_{0m}(v)&=&\nu v,~~ \nu>0,\nonumber \\
\rho_{0r}(v)&=&\lambda v^{-\beta +\frac{2}{3}},~~ \lambda >0,
\end{eqnarray}
and take the limit \eqref{eq:defx}, we arrive to algebraic equation
\begin{equation} \label{eq:algebraic}
X_0=\frac{2}{1-2\mu X_0+\frac{\lambda}{9X_0^2}} \rightarrow 2\mu X_0^3-X_0^2+2X_0-\frac{\lambda}{9}=0.
\end{equation}
This equation has at least one positive finite root in the interval $(0,1)$ if $1+2\mu-\frac{\lambda}{9}>0$.
Another important property of the solution  \eqref{eq:mass2} is that it is not asymptotically flat and has an outer apparent horizon with properties similar to the cosmological one.

\section{A model with interaction leading to regular black hole formation}
The model \eqref{eq:mass2} assumes that interaction take place everywhere inside the collapsing cloud and it is constant. In this model we assume that interaction take place at some radius inside the star and increases towards the center, i.e. 
\begin{eqnarray} \label{eq:conserv}
\rho'_dr&+&2\rho_d=\beta(r)\rho_r,\nonumber \\
\rho'_rr&+&\frac{8}{3}\rho_r=-\beta(r)\rho_r.
\end{eqnarray}
And we also assume that this interaction grows towards the centre, i.e. $\beta'<0$.
General solution of the system of differential equations \eqref{eq:conserv} is
\begin{eqnarray} \label{density_reg}
\rho_r&=&\rho_{0r}(v)e^{\int \frac{\beta(v,r)-\frac{8}{3}}{r}dr},\nonumber \\
\rho_m&=&r^{-2}\left[c_1(v)-\int r\beta(v,r)\rho_{0r}e^{\int \frac{\beta(v,r)-\frac{8}{3}}{r}dr}dr\right].
\end{eqnarray}
Where $c_1(v)$ and $\rho_{0r}(v)$ are functions of integration. To solve this system, one should use expression for energy density \eqref{eq:density}, i.e.
\begin{equation}
\rho=\rho_r+\rho_m=\frac{2M'}{r^2}.
\end{equation}
and the formal solution is
\begin{equation} \label{eq:massprom}
M(v,r)=M_0(v)+\frac{1}{2}\int r^2\left(\rho_r+\rho_m\right)dr.
\end{equation}
One can obtain regular center if the following conditions are held
\begin{itemize}
\item the densities \eqref{eq:density_reg} should be constant. Note it is possible only if the expression in squared brakets is equal to zero;
\item the limit $\lim\limits_{r\rightarrow 0}M(v,r)=0$ should be held. 
\end{itemize}
From previous section, one can see that we can satisfy the first condition but we failed to satisfy the second one. Below we give an explycit example when regular black hole might form.
\subsection{Example}
Let us consider an explycit example. We assume that interaction starts at some radius $R$ and it increases towards the center, i.e. we assume
\begin{equation}
\beta(v,r)=\frac{8}{3}-a(v)r.
\end{equation}
Substituting it into \eqref{eq:density_reg}, one obtains\footnote{For short notation, we don't write $v$ dependence of functions $c_1$ and $\rho_{0r}$}
\begin{eqnarray} \label{eq:density_example}
\rho_r&=&\rho_{0r}e^{-ar},\nonumber \\
\rho_m&=&r^{-2}\left[c_1+\rho_{0r}\left(\frac{2}{3a^2}+\frac{2r}{3a}-r^2\right)e^{-ar}\right].
\end{eqnarray}
It is obvious that $\rho_r$ is constant when $r\rightarrow 0$. However, we should demand that
\begin{equation}
c_1=-\frac{2\rho_{0r}}{3a^2},
\end{equation}
to satisfy the finite limit in the center of $\rho_m$. The total density is then given by
\begin{equation}
\rho=\rho_m+\rho_r=\rho_{0r}r^{-2}\left[-\frac{2}{3a^2}+\left(\frac{2}{3a^2}+\frac{2r}{3a}\right)e^{-ar}\right].
\end{equation}
The mass function corresponding to this energy density is
\begin{equation} \label{eq:mass3}
M(v,r)=M_0(v)-\frac{\rho_{0r}}{3a^2}r-\left(\frac{2}{3a^3}+\frac{1}{3a^2}r\right)\rho_{0r}e^{-ar}.
\end{equation}
Here we can satisfy the condition $\lim\limits_{r\rightarrow 0}M(v,r)=0$ if the following condition is held
\begin{equation}
M_0(v)=\frac{2\rho_{0r}(v)}{3a^3(v)}.
\end{equation}
Finaly, we arrive to the regular black hole
\begin{equation}
M(v,r)=M_0\left[1-\frac{ar}{2}-\left(1+\frac{ar}{2}\right)e^{-ar}\right].
\end{equation}
The Kretschmann scalar of newly obtained solution in the center behaves like
\begin{equation}
\lim\limits_{r\rightarrow 0} K=\frac{2M_0^2(v)}{3a^6}.
\end{equation}
It is necessary to note the following fact: the solution \eqref{eq:mass3} describes only the inner region, showing that as a result of this interaction no singularity formation occurs in the center of the black hole. This solution must be matched at some radius $r=R$ with the solutions either \eqref{eq:mass1} or \eqref{eq:mass2}. As a consequence, the solution \eqref{eq:mass3} cannot be used to compute black hole shadows or quasi-normal modes. For these purposes it is necessary to use the solutions \eqref{eq:mass1} or \eqref{eq:mass2}.
\section{Conclusion}
A black hole is the end state of the gravitational collapse of massive stars. However, a typical black hole contains a singularity and to avoid singularity formation we have to violate a strong energy condition that states that gravity must attract. The nature of the matter that prevents the singularity formation is still unknown. 
There are a number of models whose source is nonlinear electrodynamics. However, such black holes are unstable and by compensating the charge that prevented the formation of the singularity, we again come to singular black holes. Other candidates are the vacuum medium and the Hagedorn fluid. However, it is still not known how the ordinary matter of a star which undergoes continuous gravitational collapse is transformed into the exotic matter at the center of a black hole.
In this paper, we have constructed an idealized model of the gravitational collapse of dust and radiation, which, in the case of a special interaction, can lead to the formation of a regular center. However, the nature of this interaction remains unknown. For a more realistic model it is necessary to consider either general barotropic equations of state or to assume other matter inside the star and describe it by, say, a polytropic equation of state. Perhaps, in this case it will be possible to show that a regular center can be formed at a constant interaction. Perhaps, such interaction can be observed as a powerful energy flux from a collapsing star, just as the formation of a naked singularity might serve as the source of a gamma-ray burst~\cite{bib:joshi_burst, bib:vaidya_burst}. 
The aim of this paper was to show that the gravitational collapse of ordinary matter can lead to the formation of a regular center. However, the constructed model has one drawback: the obtained solution \eqref{eq:mass3} describes only a region of space-time in the immediate vicinity of the center, and therefore cannot be observed through, say, a black hole shadow. Nevertheless, at gravitational collapse, before formation of horizons, this region can be observed by an external observer as shown in the model \eqref{eq:mass2}.

\end{document}